\documentclass{PoS}

\usepackage{amsmath, amssymb,xcolor,epsfig}

\def\be{\begin{equation}}
\def\ee{\end{equation}}
\def\bea{\begin{eqnarray}}
\def\eea{\end{eqnarray}}

\def\mathswitch#1{\relax\ifmmode#1\else$#1$\fi}
\def\mathswitchr#1{\relax\ifmmode{\mathrm{#1}}\else$\mathrm{#1}$\fi}
\def\mathswitchit#1{\relax\ifmmode{#1}\else$#1$\fi}

\newcommand{\mr}{\mathrm}

\newcommand{\MSbar}{\mathswitch {\overline{\mr{MS}}}}

\def\citere#1{\mbox{Ref.~\cite{#1}}}
\def\citeres#1{\mbox{Refs.~\cite{#1}}}

\def\be{\beta}

\newcommand{\im}{\mathrm{i}}

\def\mathswitch#1{\relax\ifmmode#1\else$#1$\fi}
\def\mathswitchr#1{\relax\ifmmode{\mathrm{#1}}\else$\mathrm{#1}$\fi}
\def\mathswitchit#1{\relax\ifmmode{#1}\else$#1$\fi}

\newcommand{\lsim}
{\;\raisebox{-.3em}{$\stackrel{\displaystyle <}{\sim}$}\;}

\newcommand{\Ph}{\mathswitchr h}
\newcommand{\PH}{\mathswitchr H}

\newcommand{\PW}{\mathswitchr W}
\newcommand{\PZ}{\mathswitchr Z}

\newcommand{\PA}{\mathswitchr A}

\newcommand{\MW}{\mathswitch {M_\PW}}

\newcommand{\MZ}{\mathswitch {M_\PZ}}
\newcommand{\MH}{\mathswitch {M_\PH}}
\newcommand{\Mh}{\mathswitch {M_\Ph}}

\newcommand{\Prophecy}{{\sc Prophecy4f}}

\title{Electroweak corrections in the Two-Higgs-Doublet Model and a Singlet Extension of the Standard Model}

\ShortTitle{Electroweak corrections in the THDM and the SESM}

\author{Lukas Altenkamp, Michele Boggia, \speaker{Stefan Dittmaier}\\
        Albert-Ludwigs-Universit\"at Freiburg, Physikalisches Institut, 79104 Freiburg, Germany}

\author{Heidi Rzehak\\
University of Southern Denmark, CP$^3$-Origins,
Campusvej 55, DK-5230 Odense M, Denmark}

\abstract{We present the next-to-leading-order calculation of the partial decay widths of
light CP-even Higgs bosons decaying into four fermions in the Two-Higgs-Doublet Model and a
Singlet Extension of the Standard Model.
Different renormalization schemes are applied in the calculation, which is implemented
into the analysis tool \Prophecy.
Some sample results on the Higgs$\to4$ fermions decay widths
illustrate how the corrections reduce the dependence on the renormalization scale and the
choice of the scheme.}

\FullConference{Loops and Legs in Quantum Field Theory (LL2018)\\
		29 April 2018 - 04 May 2018\\
		St. Goar, Germany}

\begin{document}

\section{Introduction}

Precision Higgs physics at the LHC, among other things, requires 
precise predictions within specific Standard Model (SM) extensions, including at least
next-to-leading order (NLO) corrections in the electroweak (EW) sector. In particular, the renormalization of those extended models
deserves great care, in order to obtain a phenomenologically sound
parametrization of observables in terms of appropriate input parameters.
In this context, perturbative stability and the issue of gauge independence
play a central role. 
Here we briefly summarize recent work on the renormalization of two types of SM 
extensions: the Two-Higgs-Doublet Model (THDM) and a Singlet Extension of the SM (SESM).
Many SM extensions with non-minimal Higgs sectors contain a second Higgs doublet or a Higgs singlet, 
so that the THDM and the SESM 
can serve as low-energy effective theories for such models. 

As an application of the renormalization procedures of the THDM and SESM, we discuss
results from recent NLO
calculations for the decay of the light CP-even Higgs boson into four fermions, $\Ph\to\PW\PW/\PZ\PZ\to4f$,
as presented in more detail in \citere{Altenkamp:2017ldc,Altenkamp:2017kxk} and
\citere{Altenkamp:2018bcs}, respectively.
This class of Higgs decays is one of the best studied decay channels, 
in particular the decay into four charged leptons, 
which delivers a very clean experimental signal 
and plays an important role in the Higgs mass measurement.

\section{The Two-Higgs-Doublet Model and its renormalization}

The Higgs potential $V$ of the THDM is assumed to be
\begin{align}
V={}&
\textstyle
m^2_{11} \Phi^{\dagger}_1 \Phi_1+ m^2_{22} \Phi^\dagger_2 \Phi_2
-m^2_{12} (\Phi^\dagger_1 \Phi_2 + \Phi^\dagger_2 \Phi_1)
+ \frac{1}{2} \lambda_1 (\Phi^\dagger_1 \Phi_1)^2+\frac{1}{2} \lambda_2 (\Phi^\dagger_2 \Phi_2)^2
\nonumber\\ 
&
\textstyle
+ \lambda_3 (\Phi^\dagger_1 \Phi_1)(\Phi^\dagger_2 \Phi_2)+ \lambda_4 (\Phi^\dagger_1 \Phi_2)(\Phi^\dagger_2 \Phi_1)
+\frac{1}{2} \lambda_5\left[(\Phi^\dagger_1 \Phi_2)^2+(\Phi^\dagger_2 \Phi_1)^2\right], 
\label{eq:lambdapara}
\end{align}
where $\Phi_1$, $\Phi_2$ are the two Higgs doublets, $m^2_{11}$, $m^2_{12}$, $m^2_{22}$ the mass parameters, and $\lambda_1, \dots, \lambda_5$ the quartic Higgs couplings. The symmetry of the Higgs potential under $\Phi_1 \rightarrow -\Phi_1$ is only softly broken by non-vanishing values of 
{$m^2_{12}$ \cite{Gunion:1989we}.} In addition, we assume CP-conservation so that all parameters in the Higgs potential are real.
The two Higgs doublets can be decomposed as
\looseness -1
\begin{align}
\Phi_1=\begin{pmatrix} \phi_1^+ \\ (\eta_1+i \chi_1+v_1)/\sqrt{2}\end{pmatrix}, 
&& 
\Phi_2=\begin{pmatrix}\phi_2^+ \\ (\eta_2+i \chi_2+v_2)/\sqrt{2}\end{pmatrix},
\label{eq:decom}
\end{align}
where $v_1$, $v_2$ are the Higgs vacuum expectation values and $\phi_1^+,\phi_2^+$, $\eta_1,\eta_2$,  $\chi_1,\chi_2$ the charged, the neutral CP-even, and the neutral CP-odd fields, respectively. The fields with the same quantum numbers can mix, and the resulting mass eigenstates correspond to two CP-even Higgs bosons, $\Ph$ and $\PH$, where $\Ph$ denotes the lighter CP-even Higgs boson, one CP-odd Higgs boson $\PA_0$, two charged Higgs bosons $\PH^\pm$, and a neutral and two charged Goldstone bosons, $G_0$ and~$G^\pm$.  
\looseness -1

We replace the original set of parameters  of the Higgs and gauge sector
$m_{11}^2$, $m_{22}^2$, $m_{12}^2$, $\lambda_1$, $\lambda_2$, $\lambda_4$, $v_1$, $v_2$, $g_1 $, 
$g_2$, $\lambda_3$, $\lambda_5$
with $g_1$ and $g_2$ being the $U(1)$ and the $SU(2)$ gauge couplings, respectively, by 
$t_\Ph$, $t_\PH$, $\Mh$, $\MH$, $M_{\PA_0}$, $M_{\PH^+}$, $\MW$, $\MZ$, $e$, $\beta$, 
$\alpha$ (or $\lambda_3$), $\lambda_5$
with $t_\Ph$ and $t_\PH$ being the tadpole parameters. The masses of the CP-even, CP-odd, and charged Higgs bosons are $\Mh$, $\MH$, $M_{\PA_0}$, $M_{\PH^+}$,  the masses of the $\PW$ and the $\PZ$ boson are  $\MW$ and $\MZ$. The electric unit charge is  denoted by $e$. The parameter $\beta$ is defined via the ratio of the two Higgs vacuum expectation values, $\tan \beta = \frac{v_2}{v_1}$.  In our different renormalization schemes%
\footnote{Further renormalization schemes of the THDM are discussed in 
{\citeres{Santos:1996vt,Krause:2016oke,Denner:2017vms}.}}, 
we use either the quartic coupling $\lambda_3$ or the mixing angle of the CP-even Higgs bosons $\alpha$ as an input.

In all four renormalization schemes (see \citere{Altenkamp:2017ldc} for details), 
the Higgs- as well as the gauge-boson masses are chosen on-shell, and
the electric charge is defined via the ee$\gamma$ vertex in the Thomson limit.
The angles $\alpha$, $\beta$ (or $\lambda_3$ instead of $\alpha$) and 
the coupling $\lambda_5$ are treated as $\overline{\text{MS}}$ parameters.
The various renormalization schemes differ in the treatment of tadpole contributions:
\begin{itemize}
\setlength{\itemsep}{0em}
\item 
Variant 1:
The renormalized tadpole parameters $t_\phi^{\text{ren}}$ with $\phi = h, H$ vanish. The corresponding counterterm $\delta t_\phi$ is chosen in such a way that explicit one-loop tadpole contributions are canceled. 
However, this treatment introduces gauge dependences in the relation between bare parameters
\cite{Krause:2016oke}, and, hence, also in the relation between renormalized parameters and physical predictions.
\item 
Variant 2:
Following a procedure proposed by Fleischer and Jegerlehner (FJ)~\cite{Fleischer:1980ub},
the bare tadpole parameters  $t_\phi^{\text{bare}}$ vanish. 
Gauge dependences in the relation between bare parameters and in the relation between the renormalized 
parameters and physical predictions do not occur.
The inclusion of explicit tadpole contributions can, e.g., be avoided using the same setup as in the
``$t_\phi^{\text{ren}} = 0$"-variant if appropriate finite contributions in the \MSbar\ counterterms 
of $\alpha$ and $\beta$ are taken into account.
\end{itemize}

The following four different renormalization schemes~\cite{Altenkamp:2017ldc}
are applied:
\begin{itemize}
\setlength{\itemsep}{0em}
\item
$\MSbar(\lambda_3)$ scheme: 
$\lambda_3$ and $\beta$ are independent parameters and fixed in the 
\MSbar{} scheme, and the renormalized tadpole parameters vanish. 
The mixing angle $\alpha$ can be calculated from $\lambda_3$ and the 
other independent parameters using tree-level relations. 
The relation between independent parameters and predicted observables do not depend on a gauge parameter within the class of $R_\xi$ gauges at NLO, since $\lambda_3$ is a basic coupling in the Higgs potential and thus does not introduce gauge dependences, and since the \MSbar{} renormalization of $\beta$ is gauge-parameter independent in $R_\xi$ gauges at NLO~\cite{Krause:2016oke}.
\item
$\MSbar(\alpha)$ scheme: 
This scheme coincides with the $\MSbar(\lambda_3)$ scheme except that now $\alpha$ is chosen as independent parameter instead of $\lambda_3$.
This scheme suffers from some gauge dependence in the relation between renormalized parameters and predicted observables. Hence, for a meaningful comparison with data, all predictions using this renormalization scheme should be performed in the same gauge. 
We use the 't~Hooft--Feynman gauge.
\item
FJ$(\alpha)$ scheme: 
$\alpha$ and $\beta$ are independent parameters, and the tadpoles are treated following the gauge-independent FJ
prescription, $t_\phi^{\text{bare}} = 0 $.
Similar schemes are also described in \citeres{Krause:2016oke}, however,  the treatment of $m_{12}^2$ and $\lambda_5$ differs.
\item
FJ$(\lambda_3)$ scheme: 
$\beta$ and $\lambda_3$ are independent parameters, as in the
$\MSbar(\lambda_3)$ scheme, but the bare tadpole parameters are chosen to vanish. 
\end{itemize}

The parameters $\alpha$, $\beta$,
and the Higgs-quartic-coupling parameter $\lambda_5$ depend on a renormalization scale $\mu_\mr{r}$ in all four schemes. The $\mu_\mr{r}$~dependence of $\alpha$, $\beta$, 
and $\lambda_5$ is calculated by solving the  renormalization group equations in the four different renormalization schemes.

\section{The Singlet Extension of the SM and its renormalization}

The Higgs potential $V$ of the SESM is assumed to be
\begin{align}
V={}& \textstyle
= - \mu_2^2 \Phi^\dagger \Phi
               + \frac{\lambda_2}{4} (\Phi^\dagger \Phi)^2
               + \lambda_{12} \sigma^2 \Phi^\dagger \Phi
               - \mu_1^2 \sigma^2
               + \lambda_1 \sigma^4,
\end{align}
where $\Phi$ is the Higgs doublet, 
{$\sigma$ a real Higgs singlet field,}
$\mu_1^2$, $\mu_2^2$ the mass parameters, and $\lambda_1, \lambda_2, \lambda_{12}$ 
the quartic Higgs couplings. 
The potential is $\mathbb{Z}_2$-symmetric with respect to 
$\sigma \rightarrow -\sigma$, and all the parameters in $V$ are real quantities.
{This choice represents the most simple singlet extension,
which nevertheless bears generic features (such as Higgs mixing)
of more comprehensive versions, which, e.g., make use of charged
singlet scalars~\cite{Silveira:1985rk}.}
The Higgs fields are decomposed as
\looseness -1
\begin{align}
\label{eq:doubletSingletDef}
 \Phi =
 \begin{pmatrix}
 \phi^+ \\ \frac{1}{\sqrt{2}} (v_2 + h_2 + \im \chi)
 \end{pmatrix},
 \quad
 \sigma = v_1 + h_1,
\end{align}
where $v_1$, $v_2$ are the Higgs vacuum expectation values 
{and $\phi^+$, $\chi$}
denote the would-be Goldstone-boson fields.
The fields {$h_{1,2}$} mix to two CP-even Higgs bosons $\Ph$ and $\PH$ with masses $\Mh$ and $\MH$, 
where  $\Mh<\MH$ by definition.

We replace the original set of parameters  of the Higgs and gauge sector
$\mu_1^2$, $\mu_2^2$, $\lambda_1$, $\lambda_2$, $\lambda_{12}$,
$v_1$, $v_2$, $g_1$, $g_2$, 
by $t_\Ph$, $t_\PH$, $\Mh$, $\MH$, $\MW$, $\MZ$, $e$, 
$\alpha$, $\lambda_{12}$,
with $t_h$ and $t_H$ being again the tadpole parameters and the SM-like parameters $\MW$, $\MZ$, $e$, etc.,
are playing the same role as in the THDM 
{described} above.
In our different renormalization schemes%
\footnote{Further renormalization schemes of the SESM 
are discussed in \citeres{Denner:2017vms,Kanemura:2015fra}.}, we use 
the Higgs masses, the quartic coupling $\lambda_{12}$, and the mixing angle $\alpha$ of 
the Higgs bosons as input.

In our renormalization schemes (see \citere{Altenkamp:2018bcs} for details), 
{all particle masses} are chosen on-shell, 
the electric charge is defined in the Thomson limit,
while the angle $\alpha$ and 
the coupling $\lambda_{12}$ are treated as $\overline{\text{MS}}$ parameters.
As in the THDM, we employ the two different variants of tadpole treatments
described in the previous section, called $\MSbar$ and FJ in the following.
We recall that among those two schemes only the FJ scheme delivers are
gauge-independent parametrization of amplitudes in terms of the input parameters,
while all results in the $\MSbar$ scheme are tied to a specific gauge,
which is taken as the 't~Hooft--Feynman gauge.

\section{Numerical results for the partial decay width for $\Ph\to\PW\PW/\PZ\PZ\to4f$}

\subsection{Outline of the calculation}

The computer program
\Prophecy~\cite{Bredenstein:2006rh}
provides a ``\textbf{PROP}er description of the \textbf{H}iggs d\textbf{EC}a\textbf{Y} into \textbf{4 F}ermions'' and calculates 
observables for the decay process $\Ph {\to} \PW\PW/\PZ\PZ {\to} 4 f$ at NLO EW+QCD in  the SM. 
We have extended \Prophecy\ implementing the corresponding decay of the light, neutral
CP-even Higgs boson of the THDM in such a way that the features of 
\Prophecy\ and its applicability basically remain the same. 
We have performed two independent calculations and implementations, as described in 
\citeres{Altenkamp:2017ldc,Altenkamp:2017kxk} in detail:
\begin{itemize}
\setlength{\itemsep}{0em}
\item 
For one calculation, 
we have used a model file generated by {\tt  FeynRules} \cite{Christensen:2008py}, and for the other one an inhouse model file.
The amplitudes for the virtual corrections have been generated with two different versions of 
{\tt  FeynArts}~\cite{Kublbeck:1990xc} and algebraically reduced
with {\tt  FormCalc}~\cite{Hahn:1998yk} in the first calculation and with inhouse 
Mathematica routines {in the second.}
\looseness-1
  
\item 
Masses of final-state fermions are neglected, 
but masses are taken into account in closed fermion loops. 
Hence, the contribution of diagrams with a closed fermion loop coupling to the 
Higgs boson does not vanish. 
We have implemented four different THDM types (Type~1, Type~2, "flipped", "lepton-specific") 
that differ in how the down-type quarks and charged leptons couple to the two Higgs doublets. 
Since the up-type quarks couple always in the same manner in all of the four types of THDM and since the 
dominating contribution originates from the 
top-quark loop, the differences between the types are negligible.
  
\item
Infrared divergences are treated applying dipole 
subtraction~\cite{Catani:1996vz}.
\item 
The $\PW$ and $\PZ$ resonances are treated in the  complex-mass scheme as 
{described} in \citere{Denner:2005fg}.
The evaluation of loop integrals is performed with the {\tt  Collier} library~\cite{Denner:2016kdg}.
\end{itemize}

\subsection{Numerical results for $\Ph\to\PW\PW/\PZ\PZ\to4f$ in the THDM}

In this section, we show some sample results for the partial decay width 
$\Gamma^{\Ph\to4f}_\mr{THDM}$
for $\Ph\to\PW\PW/\PZ\PZ\to4f$ for a scenario (scenario~A) inspired by  Ref.~\cite{Haber:2015pua} for the Type~1 THDM:
\begin{equation}
\Mh = 125 \text{ GeV}, \quad \MH = 300 \text{ GeV},\quad M_{\PA_0} = M_{\PH^+} = 460 \text{ GeV},\quad \lambda_5 = -1.9, \quad \tan\beta= 2.
\end{equation}
Within our calculation, we choose the central renormalization scale as the average mass  of all scalar degrees of freedom, $\mu_0 = (\Mh + \MH + M_{\PA_0} + 2 M_{\PH^+})/5$.
 
\begin{figure}[b]
\centerline{
\includegraphics[scale=1.]{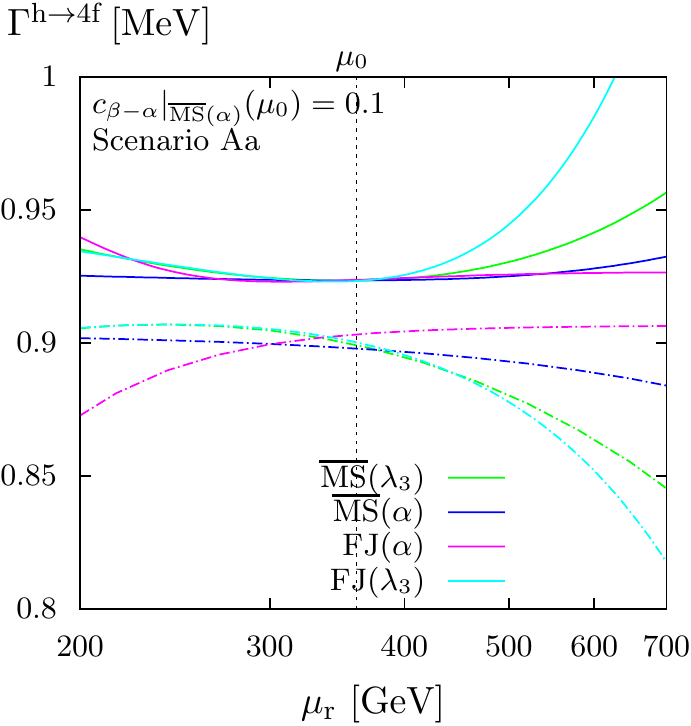} \qquad
\includegraphics[scale=1.]{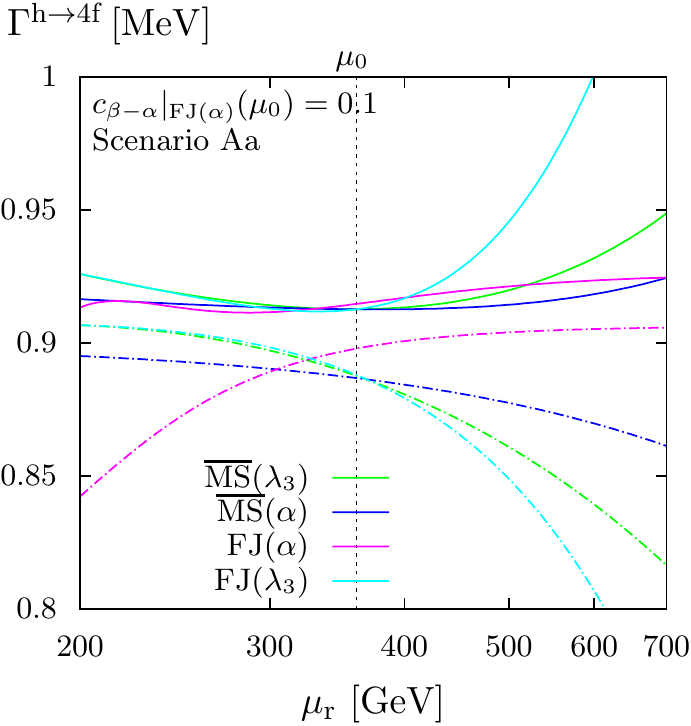}
}
\vspace*{-.5em}
\caption{The renormalization scale dependence of $\Gamma^{\Ph\to4f}_\mr{THDM}$
for two of the four different input schemes: $\MSbar(\alpha)$ (left) and FJ$(\alpha)$ (right). 
In each plot, the parameters are converted 
to the other schemes,  $\MSbar(\lambda_3)$ (green), $\MSbar(\alpha)$ (blue), FJ$(\alpha)$ (magenta), and FJ$(\lambda_3)$ (turquoise). 
The solid lines include NLO EW corrections, the dashed ones show the 
LO result. 
{(Figure taken from \protect \citere{Altenkamp:2017kxk}.)} \label{fig:muscan_thdm}}
\end{figure}

In Fig.~\ref{fig:muscan_thdm},  the renormalization scale dependence of $\Gamma^{\Ph\to4f}_\mr{THDM}$, which is obtained by summing over all partial widths of the h~boson with massless $4f$~final states, is shown. We fix $\cos (\beta - \alpha) = c_{\beta-\alpha} = 0.1$ (scenario Aa). 
The two plots correspond to the input parameters given in
the $\MSbar(\alpha)$ and the FJ$(\alpha)$
renormalization schemes. The dashed curves represent the leading-order (LO) results, however, it should be noted that the input parameters have been converted to the respective scheme denoted by the different line colours. Hence, the strict LO result is only represented by the line corresponding to the input scheme, i.e.\ for example, in the left plot, the strict LO curve is given by the dark blue $\MSbar(\alpha)$ line. The differences between the dashed lines at the scale $\mu_0$ are only due to conversion effects, while at the other scales also the different running behaviour of the $\MSbar$ parameters in the different schemes plays a role. 
Note that it is important to specify not only the parameter values of a certain 
scenario, but also the renormalization scheme, in which these parameters are to be interpreted.
The solid lines show the NLO result including only the EW corrections. A clear plateau around the central renormalization scale $\mu_0$ is 
visible, and there is a clear reduction on the scale dependence going from LO to NLO. 

A detailed discussion of further results, including also differential distributions and
more delicate THDM scenarios, can be
found in \citeres{Altenkamp:2017ldc,Altenkamp:2017kxk}.

\subsection{Numerical results for $\Ph\to\PW\PW/\PZ\PZ\to4f$ in the SESM}

In this section, we show some sample results for the partial decay width 
$\Gamma^{\Ph\to4f}_\mr{SESM}$
for $\Ph\to\PW\PW/\PZ\PZ\to4f$ for a scenario (BHM200$^+$) 
inspired by  Ref.~\cite{Robens:2016xkb} for the SESM:
\begin{equation}
\Mh = 125 \text{ GeV}, \quad \MH = 200 \text{ GeV},\quad s_\alpha=0.29,\quad \lambda_{12} = 0.07.
\end{equation}
As central renormalization scale we choose $\mu_0 = \Mh$.
 
\begin{figure}[b]
\centerline{
\includegraphics[scale=1.]{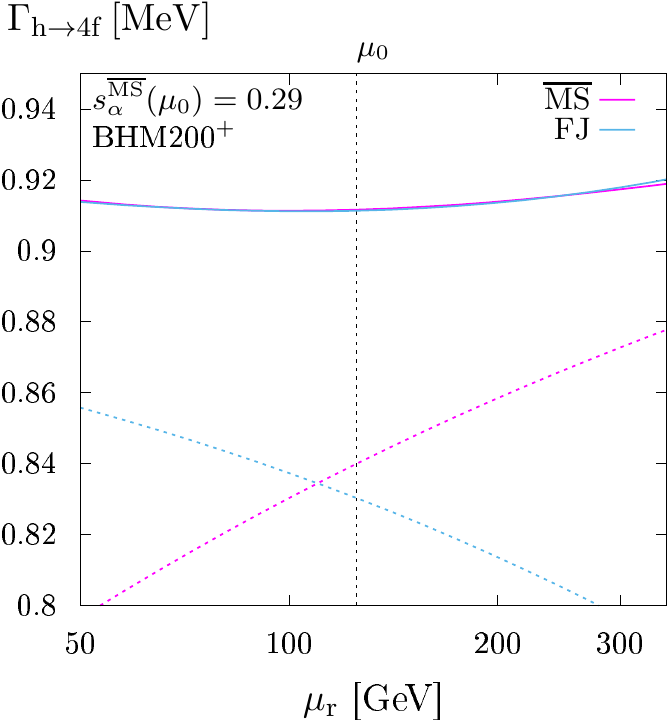}
\qquad
\includegraphics[scale=1.]{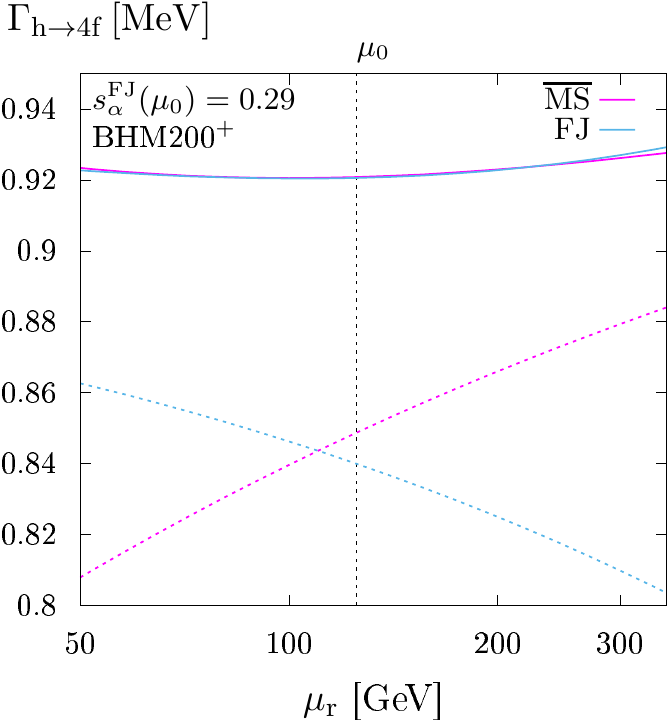}
}
\vspace*{-.5em}
\caption{The renormalization scale dependence of $\Gamma^{\Ph\to4f}_\mr{SESM}$
for {two input schemes:} $\MSbar$ (left) and FJ (right). 
In each plot, the parameters are converted 
to the other schemes,  $\MSbar$ (magenta) and FJ (turquoise). 
The solid lines include NLO QCD+EW corrections, the dashed ones show the 
LO result. {(Figure taken from \protect \citere{Altenkamp:2018bcs}.})}
\label{fig:muscan_sesm}
\end{figure}

In Fig.~\ref{fig:muscan_sesm}, the renormalization scale dependence of $\Gamma^{\Ph\to4f}_\mr{SESM}$ is shown.
The two plots correspond to the input parameters given in the $\MSbar$ and FJ scheme, respectively.
The dashed curves represent the LO results, and 
the input parameters have been converted to the respective scheme denoted by the different line colours. 
Hence, the strict LO result is only represented by the line corresponding to the input scheme, 
i.e.\ for example, in the left plot, the strict LO curve is given by the 
{magenta} $\MSbar$ line. 
The differences between the dashed lines at the scale $\mu_0$ are only due to conversion effects, 
while at the other scales also the different running behaviour of the $\MSbar$ parameters 
$\alpha$ and $\lambda_{12}$ in the different schemes plays a role. 
The solid lines show the NLO result including QCD+EW corrections. 
A very broad plateau around the central renormalization scale $\mu_0$ is 
visible, with a drastic reduction on the scale dependence going from LO to NLO. 
The residual scale dependence and the scheme dependence are
generically (i.e.\ also for typical scenarios
with larger $\MH$) reduced from the few-percent level at LO to less than $\lsim0.3\%$ at NLO,
which is covered by the size of missing higher-order corrections of NLO predictions.

A detailed discussion of further results, including differential distributions, can be
found in \citere{Altenkamp:2018bcs}.

\section{Conclusions}

We have briefly summarized proposals for
the renormalization of the THDM and an SESM, a subject that
deserves great care in order to obtain a phenomenologically sound
parametrization of observables in terms of appropriate input parameters.
In this context, perturbative stability and the issue of gauge independence
play a central role. 
The proposed renormalization schemes
are based on on-shell renormalization conditions for particles masses
and the electromagnetic coupling and on $\MSbar$ conditions for the non-standard
parameters that are not directly related to observables, such as 
mixing angles or scalar self-couplings. The use of those $\MSbar$ conditions 
leads to a dependence of the renormalization schemes on the treatment of
tadpoles, i.e.\ on the details how the vacuum state is defined. We employ
and compare results from two different tadpole schemes.

{As an application} of the described schemes, we have calculated the
NLO corrections to all decays of the light CP-even Higgs bosons of the THDM and the SESM
into four fermions, $\Ph\to\PW\PW/\PZ\PZ\to4f$. 
We have shown typical 
{results of the two models} with moderately heavy Higgs-boson
masses and mixing angles near the alignment regions.
For those cases, we find textbook-like reductions of the perturbative uncertainties
(dependence on the renormalization scale and schemes)
in the transition from LO to NLO.
In this context we emphasize that 
it is generally important to specify not only the parameter values of a certain 
scenario, but also the renormalization scheme, in which these parameters are to be interpreted.
For a meaningful comparison of results in different schemes, the parameters have to be
properly translated between the schemes.
A detailed discussion of further results in the THDM and the SESM,
covering more scenarios and corrections to differential distributions, 
can be found in \citeres{Altenkamp:2017ldc,Altenkamp:2017kxk} and \citere{Altenkamp:2018bcs},
respectively. 
While the THDM renormalization schemes start to show problems with perturbative
stability in specific parameter regions (mass degeneracy of Higgs bosons, large
mass gaps between Higgs bosons, extreme mixing, etc.),
the proposed schemes for the SESM behave rather robust.
Generically, we find that the shapes of differential distributions 
for $\Ph\to4f$ will not help to
tell the THDM and SESM from the SM, since the corrections beyond the SM do not lead
to further distortions on the distributions.
\looseness -1

The extended version of \Prophecy,
which covers the extensions to the THDM and the SESM,
will be available from its hepforge 
webpage\footnote{\tt http://prophecy4f.hepforge.org/index.html} soon.

\end{document}